\documentclass[twoside]{article}
\usepackage{fleqn,espcrc2}

\usepackage{epsf}
\newcommand{\AmS}{{\protect\the\textfont2
  A\kern-.1667em\lower.5ex\hbox{M}\kern-.125emS}}
\def\l{\left}
\def\r{\right}

\def\Tr{\,\mathrm{Tr}}

\def\ev{\,\mathrm{e\kern-0.1em V}}
\def\kev{\,\mathrm{ke\kern-0.1em V}}
\def\mev{\,\mathrm{Me\kern-0.1em V}}
\def\gev{\,\mathrm{Ge\kern-0.1em V}}
\def\tev{\,\mathrm{Te\kern-0.1em V}}

\def\n#1e#2n{{#1}\times 10^{#2}}

\def\la{\langle}
\def\ra{\rangle}

\def\n{\linebreak}

\def\ln#1{\mathrm{ln}\l(#1\r)}

\def\bea{\begin{eqnarray}}
\def\eea{\end{eqnarray}}
\def\beq{\begin{equation}}
\def\eeq{\end{equation}}

\def\Tr{\mathrm{Tr}}

\def\eq#1{Eq.~(\ref{#1})}
\def\eqs#1#2{Eqs.~(\ref{#1}) and (\ref{#2})}

\def\fig#1{Fig.~\ref{#1}}

\def\gev{\mbox{ GeV}}
\def\mev{\mbox{ MeV}}

\title{Chiral symmetry breaking from Ginsparg-Wilson fermions
\thanks{Presented by L.\ Lellouch at Lattice '99. See 
also \cite{Hernandez:1999cu}.}}

\author{Pilar Hern\'andez$^{\mathrm a}$\thanks{On leave from
    Departamento de F\'{\i}sica Te\'orica, Universidad de Valencia.},
  Karl Jansen$^{\mathrm a}$\thanks{Heisenberg Foundation Fellow.} and
  Laurent Lellouch\address{CERN, 1211 Geneva 23,
    Switzerland}\thanks{On leave from CPT
    CNRS Luminy, F-13288 Marseille Cedex 9, France. Work supported in
    part by TMR, EC-Contract No.  ERBFMRX-CT980169~(EURODA$\Phi$NE).}}

\begin{document}

\begin{abstract}
  
  We calculate the large-volume and small-mass dependences of the
  quark condensate in quenched QCD using Neuberger's operator. We find
  good agreement with the predictions of quenched chiral perturbation
  theory, enabling a determination of the chiral lagrangian parameter
  $\Sigma$, up to a multiplicative renormalization.

%\vspace{1pc}

\end{abstract}

% typeset front matter (including abstract)
\maketitle

\section{Introduction}

Spontaneous chiral symmetry breaking (S$\chi$SB) is fundamental to our
understanding of low energy hadronic phenomena and it is thus
important to demonstrate quantitatively that it is a consequence of
QCD.  A natural candidate for such investigations is the numerical
simulation of QCD on a spacetime lattice. S$\chi$SB, however, presents
the lattice approach with a twofold challenge.

The first is that spontaneous symmetry breaking does not occur in a
finite volume. In QCD, a possible signal of S$\chi$SB is the presence
of a non-vanishing quark condensate defined as:
\beq
-\Sigma\equiv\la\bar qq\ra =\lim_{\mbox{\tiny $m\to 0$}}
\lim_{\mbox{\tiny $V\to\infty$}} \la\bar
qq\ra_{m,V}
\ ,
\label{eq:conddef}
\eeq
where $\la\bar qq\ra_{m,V}$ is the condensate for finite volume $V$
and mass $m$.  The double limit in \eq{eq:conddef} is rather
challenging numerically!  To get around this problem, we resort to a
finite-size scaling analysis.  This involves studying the scaling of
the condensate with $V$ and $m$ as the limit of restoration of $\chi$S
is approached ($m\to 0$, $V$ finite).

Such a study requires good control over the chiral properties of the
theory, which is the second challenge. Indeed, at finite lattice
spacing, ``reasonable'' discretizations of fermions either break
continuum $\chi$S explicitly or lead to extraneous fermion species
\cite{Nielsen:1981hk}. To minimize this problem, we resort to recently
rediscovered \cite{Hasenfratz:1997ft,Neuberger:1998wv} Ginsparg-Wilson
(GW) fermions \cite{Ginsparg:1982bj} which break continuum $\chi$S in
a very mild and controlled fashion and actually have a slightly
generalized $\chi$S even at finite lattice spacing
\cite{Luscher:1998pq}.

\vskip -9.5cm
\rightline{CERN-TH/99-273}
\rightline{CPT-99/PE.3886}
\vskip +9.1cm

%\vspace{-0.2cm}

\section{Light quarks on a torus}

In a large periodic box of volume $V=L^4$ such that $F_\pi L\gg 1$,
for small quark masses and assuming the standard pattern of S$\chi$SB
with $N_f\ge 2$, the QCD partition function is dominated by the
nearly massless pions; the system can be described with the first
few terms of a chiral lagrangian \cite{Gasser:1988zq}. If, in addition,
$m\to 0$~\footnote{We assume here for simplicity that the $N_f$
flavors all have mass $m$.} so that $M_\pi
L\simeq\frac{\sqrt{2m\Sigma}}{F_\pi}L\ll 1$, the global mode of
the chiral lagrangian field $U\in SU(N_f)$ dominates the partition
function, leading to a regime of restoration of
$\chi$S \cite{Gasser:1987ah}.

In the quenched approximation to which we restrict here, topological
zero modes of the Dirac operator induce $1/m$ singularities in
$\la\bar qq\ra_{m,V}$ as $m\to 0$. To subtract these contributions, we
work in sectors of fixed topological charge.  Generalizing the line of
argument given above, the partition function
$Z_\nu$, in a sector of topological charge $\nu$, was recently
evaluated \cite{Osborn:1998qb} for the quenched case\footnote{The
  original unquenched treatment is given in \cite{Leutwyler:1992yt}.}.
The quark condensate in sector $\nu$, proportional to the derivative
of $\ln {Z_\nu}$ w.r.t.\ $m$, is then $-\Sigma_\nu\equiv\la\bar
qq\ra_{m,V,\nu}$, such that \cite{Osborn:1998qb}
\beq
\frac{\Sigma_\nu}{\Sigma} = z \; [ I_\nu(z) K_\nu(z) + I_{\nu+1}(z)
K_{\nu-1}(z)]+\frac{\nu}{z},
\label{qpt}
\eeq
where $z \equiv m \Sigma V$ and $I_\nu(z)$, $K_\nu(z)$ are the
modified Bessel functions. As advertised, there is a divergence $\sim
1/m$ in sectors with topology. These terms, however, are independent
of $\Sigma$.

\eq{qpt} summarizes the scaling of the quark condensate with the
volume and quark mass in the global mode regime, as a function
of only one non-perturbative parameter: $\Sigma$. Thus, by fitting the
dependence of the finite-volume condensate in quark mass and volume to
Monte Carlo data, we can extract $\Sigma$ in a perfectly controlled
manner.

\section{Ginsparg-Wilson fermions}

To perform the finite-volume scaling analysis outlined above, we need
to be able to reach the chiral restoration regime without excessive
fine-tuning and we need an index theorem to control the contribution
of topological zero modes. Both these requirements are satisfied by
GW fermions \cite{Hasenfratz:1998jp}. In particular, 
the leading cubic UV divergence of the condensate is known
analytically for GW fermions and can thus be subtracted {\rm
exactly}. The resulting subtracted condensate,
$\Sigma^{sub}_{\nu}$, however, is still divergent:
\begin{eqnarray}
\Sigma^{sub}_{\nu}(a) = C_2 \frac{m}{a^2} + \cdots +
\Sigma_{\nu}, \label{eq:sigsub}
\end{eqnarray}
where $a$ is the lattice spacing.  The coefficients of the divergences
are not known a priori and have to be determined, preferably
non-perturbatively. For the values of $m$ and $a$ considered below,
however, only the quadratic divergence is important numerically,
weaker divergences being suppressed by higher powers of $m$. A final
multiplicative renormalization is still required to eliminate a
residual logarithmic UV divergence in $\Sigma_{\nu}$.

In the present work, we use Neuberger's implementation of
GW fermions encoded in the Dirac
operator \cite{Neuberger:1997fp,Neuberger:1998wv}:
\beq
aD_N=(1+s)\left[
1-A/\sqrt{A^{\dagger}A}\right]
\ ,\label{eq:Dndef}
\eeq
with $A=1+s - aD_W$ where $D_W$ is the standard Wilson-Dirac operator. The
parameter $s$ must satisfy $|s|<1$.

\section{Numerical results}

We work in the quenched approximation on hypercubic lattices with
periodic boundary conditions for gauge and fermion
fields. We choose $\beta=5.85$, which corresponds to $a^{-1}\simeq
1.5\gev$ \cite{Edwards:1997xf}, and use standard methods to obtain
decorrelated gauge-field configurations.

To evaluate $1/\sqrt{A^\dagger A}$ in \eq{eq:Dndef}, we use a
Chebyshev approximation, $P_{n,\epsilon}(A^\dagger A)$, where
$P_{n,\epsilon}$ is a polynomial of degree $n$, which gives an
exponentially converging approximation to $1/\sqrt{x}$ for $x\in
[\epsilon,1]$ \cite{Hernandez:1998et}. The cost of a multiplication by
$D_N$ is linear in $n$ and becomes rapidly high. To reduce $n$
substantially, we perform the improvements described in
\cite{Hernandez:1999cu}. We take $s=0.6$, a value at which
Neuberger's operator is nearly optimally local for $\beta=5.85$
\cite{Hernandez:1998et}.

To determine whether a gauge configuration belongs to the $\nu=0$ or
$\pm 1$ sectors, we compute the few lowest eigenvalues of $D_N^\dagger
D_N$ by minimizing the relevant Ritz functional \cite{ritz}. As pointed
out in \cite{Edwards:1998wx}, it is advantageous for this computation,
as well as for the inversion of $(D_N^\dagger+m)(D_N+m)$, to stay in a
given chiral subspace. Having determined the topological charge of a
configuration, we then obtain the condensate of \eq{eq:sigsub} in
three volumes ($8^4$, $10^4$ and $12^4$) by computing
\beq
\Sigma^{sub}_\nu= \frac{1}{V}\l\langle\Tr'\l\{
\frac{1}{D_N+m}+{\mathrm h.c.}-\frac{a}{1+s}\r\}\r\rangle_\nu, 
\label{eq:condtrace}
\eeq
where the trace is taken in the chiral sector opposite to that
with the zero modes \cite{Edwards:1998wx} and the gauge average is
performed in a sector of fixed topology $\nu$. With this definition, terms
$\sim 1/m$ in \eq{qpt} are absent\footnote{Though not shown
  explicitly in \protect\eq{eq:condtrace}, we correctly account for
  the real eigenvalues of $D_N$ at $2(1+s)/a$.}. Three gaussian
sources and a multimass solver \cite{Jegerlehner:1997rn} were used
to compute the trace in \eq{eq:condtrace} for seven values of $m$.

\begin{figure}[htb]
%\vspace{9pt}
%%BoundingBox: 18 144 592 718
\epsfxsize=7cm\epsffile[18 270 480 580]{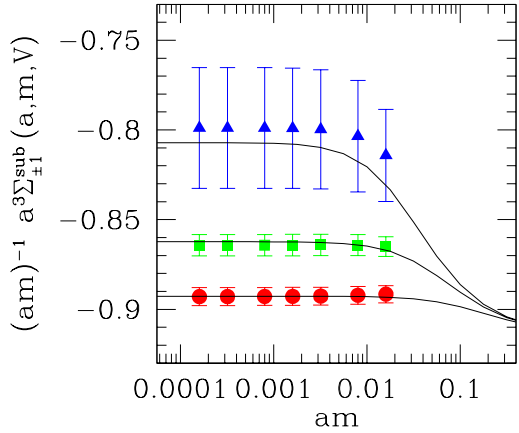} \caption{\small\it
Mass dependence of the condensate for the $8^4$ (circles),
$10^4$ (squares) and $12^4$ (triangles) lattices. The
curves result from a fit to \protect\eqs{eq:sigsub}{qpt}.
\vspace{-0.5cm}}
\label{fig:fss}
\end{figure}

We show in \fig{fig:fss} our results for $a^3\Sigma^{sub}_{\nu=\pm
  1}/am$ as a function of bare quark mass. We have 15, 10 and 7 gauge
configurations on our $8^4$, $10^4$ and $12^4$ lattices,
respectively\footnote{For the larger volumes, $\nu=0$ configurations
  are rare while the calculation of $\Sigma^{sub}_{\nu=0}$ requires
  large statistics \cite{Hernandez:1999cu}. $|\nu|>1$ configurations,
  on the other hand, are rare in the smaller volumes.}. The solid
lines are a fit of the data to \eqs{eq:sigsub}{qpt} for all volumes
and masses.  This fit has only two parameters, namely $\Sigma$ and the
coefficient of the quadratic divergence.  We find
$a^3\Sigma=0.0032(4)$ and $C_2=-0.914(8)$.

Clearly, the formulae derived in quenched $\chi$PT give a very good
description of the numerical data. The value of $\Sigma$ that we
extract is, in physical units, $\Sigma(\mu\sim
1.5\gev)=(221^{+8}_{-9}\mev)^3$, up to a multiplicative
renormalization constant, which has not been computed yet for
Neuberger's operator. The quoted error on $\Sigma$ is purely
statistical and the statistics are rather small. Quenching and
discretization errors, for instance, as well as possible contributions
from higher orders in $\chi$PT are not accounted for. Nevertheless,
the value obtained and the agreement with q$\chi$PT support the
standard scenario of S$\chi$SB.

We further consider the mean value of the lowest non-zero eigenvalue
of $\sqrt{D^\dagger_N D_N}$ in different topological sectors.  In
Random Matrix Theory the distributions of these eigenvalues are given
solely in terms of $\Sigma$ \cite{lmin}. Our determination of $\Sigma$
therefore yields predictions for the mean values. These can then be
compared to the average values obtained in simulation. With our $8^4$
results, we find agreement within roughly one standard deviation for
$|\nu|=1$ (29 configurations) and two for $\nu=0$ (41 configurations)
\cite{Hernandez:1999cu}.

{\small\it Note: Related work with Neuberger's operator can be found
  in \cite{Edwards:1998wx,Edwards:1999ra,Damgaard:1999tk}.}


\begin{thebibliography}{9}

\bibitem{Hernandez:1999cu}
P.~Hern\'andez, K.~Jansen and L.~Lellouch,
%``Finite-size scaling of the quark condensate in quenched lattice QCD,''
\texttt{hep-lat/9907022}.
%%CITATION = HEP-LAT 9907022;%%


\bibitem{Nielsen:1981hk}
H.B.~Nielsen and M.~Ninomiya,
%``No Go Theorem For Regularizing Chiral Fermions,''
Phys.\ Lett.\ {\bf 105B} (1981) 219.
%%CITATION = PHLTA,105B,219;%%


\bibitem{Hasenfratz:1997ft}
P.~Hasenfratz,
%``Prospects for perfect actions,''
Nucl.\ Phys.\ {\bf PS63} (1998) 53.
%hep-lat/9709110.
%%CITATION = NUPHZ,63,53;%%


\bibitem{Neuberger:1998wv}
H.~Neuberger,
%``More about exactly massless quarks on the lattice,''
Phys.\ Lett.\ {\bf B427} (1998) 353
%hep-lat/9801031.
%%CITATION = PHLTA,B427,353;%%


\bibitem{Ginsparg:1982bj}
P.H.~Ginsparg and K.G.~Wilson,
%``A Remnant Of Chiral Symmetry On The Lattice,''
Phys.\ Rev.\ {\bf D25} (1982) 2649.
%%CITATION = PHRVA,D25,2649;%%


\bibitem{Luscher:1998pq}
M.~L\"uscher,
%``Exact chiral symmetry on the lattice and the Ginsparg-Wilson relation,''
Phys.\ Lett.\ {\bf B428} (1998) 342.
%hep-lat/9802011.
%%CITATION = PHLTA,B428,342;%%


\bibitem{Gasser:1988zq}
J.~Gasser and H.~Leutwyler,
%``Spontaneously Broken Symmetries: Effective Lagrangians At Finite Volume,''
Nucl.\ Phys.\ {\bf B307} (1988) 763.
%%CITATION = NUPHA,B307,763;%%


\bibitem{Gasser:1987ah}
J.~Gasser and H.~Leutwyler,
%``Thermodynamics Of Chiral Symmetry,''
Phys.\ Lett.\ {\bf 188B} (1987) 477.
%%CITATION = PHLTA,188B,477;%%


\bibitem{Osborn:1998qb}
J.C.~Osborn {\it et al.},
%, D.~Toublan and J.J.~Verbaarschot,
%``From chiral random matrix theory to chiral perturbation theory,''
Nucl.\ Phys.\ {\bf B540} (1999) 317;
%hep-th/9806110.
%%CITATION = NUPHA,B540,317;%%
%\bibitem{Damgaard:1998xy}
P.H.~Damgaard {\it et al.},
%, J.C.~Osborn, D.~Toublan and J.J.~Verbaarschot,
%``The microscopic spectral density of the QCD Dirac operator,''
Nucl.\ Phys.\ {\bf B547} (1999) 305.
%hep-th/9811212.
%%CITATION = NUPHA,B547,305;%%


\bibitem{Leutwyler:1992yt}
H.~Leutwyler and A.~Smilga,
%``Spectrum of Dirac operator and role of winding number in QCD,''
Phys.\ Rev.\ {\bf D46} (1992) 5607.
%%CITATION = PHRVA,D46,5607;%%


\bibitem{Hasenfratz:1998jp}
P.~Hasenfratz,
%``Lattice QCD without tuning, mixing and current renormalization,''
Nucl.\ Phys.\ {\bf B525} (1998) 401;
%hep-lat/9802007.
%%CITATION = NUPHA,B525,401;%%
%\bibitem{Hasenfratz:1998ri}
P.~Hasenfratz, V.~Laliena and F.~Niedermayer,
%``The index theorem in QCD with a finite cut-off,''
Phys.\ Lett.\ {\bf B427} (1998) 125.
%hep-lat/9801021.
%%CITATION = PHLTA,B427,125;%%


\bibitem{Neuberger:1997fp}
H.~Neuberger,
%``Exactly massless quarks on the lattice,''
Phys.\ Lett.\ {\bf B417} (1998) 141.
%hep-lat/9707022.
%%CITATION = PHLTA,B417,141;%%


\bibitem{Hernandez:1998et}
P.~Hern\'andez, K.~Jansen and M.~L\"uscher,
%``Locality properties of Neuberger's lattice Dirac operator,''
Nucl.\ Phys.\ {\bf B552} (1999) 363.
%hep-lat/9808010.
%%CITATION = NUPHA,B552,363;%%


\bibitem{Edwards:1997xf}
R.G.~Edwards, U.M.~Heller and T.R.~Klassen,
%``Accurate scale determinations for the Wilson gauge action,''
Nucl.\ Phys.\ {\bf B517} (1998) 377.
%hep-lat/9711003.
%%CITATION = NUPHA,B517,377;%%


\bibitem{Edwards:1998wx}
%R.G.~Edwards, U.M.~Heller and R.~Narayanan,
R.G.~Edwards {\it et al.},
%``A study of chiral symmetry in quenched QCD using the overlap-Dirac  operator,''
Phys.\ Rev.\ {\bf D59} (1999) 094510.
%hep-lat/9811030.
%%CITATION = PHRVA,D59,094510;%%


\bibitem{ritz} 
%B. Bunk, K. Jansen, M. L\"uscher and H. Simma, DESY
%report (September 1994); T. Kalkreuter and H. Simma,
%Comp. Phys. Commun. {\bf 93} (1996) 33.  
B. Bunk {\it et al.}, DESY
report (Sept. 1994); T. Kalkreuter and H. Simma,
Comput. Phys. Commun. {\bf 93} (1996) 33.  


\bibitem{Jegerlehner:1997rn}
B.~Jegerlehner,
%``Multiple mass solvers,''
Nucl.\ Phys.\ {\bf PS63} (1998) 958.
%hep-lat/9708029.
%%CITATION = NUPHZ,63,958;%%


\bibitem{lmin} P.J. Forrester, Nucl. Phys. {\bf B402} (1993) 709;
%\bibitem{Nishigaki:1998is}
%S.M.~Nishigaki, P.H.~Damgaard and T.~Wettig,
S.M.~Nishigaki {\it et al.},
%``Smallest Dirac eigenvalue distribution from random matrix theory,''
Phys.\ Rev.\ {\bf D58} (1998) 087704.
%hep-th/9803007.
%%CITATION = PHRVA,D58,087704;%%


\bibitem{Edwards:1999ra}
%R.G.~Edwards, U.M.~Heller, J.~Kiskis and R.~Narayanan,
R.G.~Edwards {\it et al.},
%``Quark spectra, topology and random matrix theory,''
Phys.\ Rev.\ Lett.\ {\bf 82} (1999) 4188.
%hep-th/9902117.
%%CITATION = PRLTA,82,4188;%%


\bibitem{Damgaard:1999tk}
P.H.~Damgaard, R.G.~Edwards, U.M.~Heller and R.~Narayanan,
%``Universal scaling of the chiral condensate in finite-volume gauge  theories,''
\texttt{hep-lat/9907016}.
%%CITATION = HEP-LAT 9907016;%%















\end{thebibliography}
\end{document}